\documentclass[a4paper]{jpconf}
\usepackage{graphicx}
\usepackage[hidelinks]{hyperref}
\usepackage{color}
\usepackage{amsmath,amsfonts,amssymb,bm}

\usepackage{lipsum}

\usepackage{fancyhdr}
\def\beq{\begin{equation}}
\def\eeq{\end{equation}}
\def\beqn{\begin{eqnarray}}
\def\eeqn{\end{eqnarray}}

\def\r {{\bf r}}
\def\p {{\bf p}}

\def\P {{\bf P}}

\def\bi {\mbox{\boldmath $i$}}

\def\r {{\bf r}}

\def\p {{\bf p}}

\def\P {{\bf P}}

\begin{document}

\title{Condensates Breaking Up Under Rotation}

\author{S~Dutta$^{1,2}$, A~U~J~Lode$^3$ and O~E~Alon$^{1,2}$}
\address{$^{1}$ Department of Physics, University of Haifa, 3498838 Haifa, Israel}
\address{$^{2}$ Haifa Research Center for Theoretical Physics and Astrophysics, University of Haifa, 3498838 Haifa, Israel}
\address{$^{3}$ Institute of Physics, Albert-Ludwig University of Freiburg, Hermann-Herder-Strasse 3, 79104 Freiburg, Germany}

\ead{sdutta@campus.haifa.ac.il}

\begin{abstract}
The ground state of a rotating
Bose-Einstein condensate trapped in a two-dimensional anharmonic--anisotropic potential
is analyzed numerically at the limit of an infinite number of particles.
We find that the density
breaks up along the $x$ direction in position space and
along the $p_y$ direction in momentum space together with the acquisition of angular momentum.
Side by side,
the anisotropies of the many-particle position variances along the $x$ and $y$ directions
and of the many-particle momentum variances along the $p_y$ and $p_x$
directions become opposite when computed at the many-body and mean-field levels of theory. 
All in all,
the rotating bosons are found to possess
unique correlations 
at the limit of an infinite number of particles,
both in position and momentum spaces,
although their many-body and mean-field energies per particle and densities per particle coincide
and the condensate fraction is 100\%.
Implications are briefly discussed.
\end{abstract}

\section{Introduction}\label{Intro}

Rotating Bose-Einstein condensates have amply been researched
when the emergence of quantum vortex states and
their properties are put up front, see, e.g.,
Refs.~[1-16].
Here, we examine a different
phenomenon,
the breakup of a condensate held in an anharmonic--anisotropic trap potential upon rotation.
Building on recent investigations made both at the mean-field level \cite{BUPMF}
and particularly at the many-body level of theory for finite systems \cite{BUPMB},
we report on unique correlations in the ground state that survive the limit of an infinite number of particles,
both in position and momentum spaces,
of condensates breaking up upon rotation.

It is well known that at the infinite-particle-number limit, i.e.,
when the number of bosons is increased to infinity while holding the interaction parameter (the product of the number of particles times the interaction strength) fixed,
the many-body energy per particle,
density per particle, and 100\% condensate fraction boil down to the mean-field results,
also for rotating bosons
[19-21].
On the other hand, variances of observables,
such as the position, momentum, and angular-momentum many-particle operators,
can be quite different when comparing predictions made at the many-body and mean-field levels of theory 
for finite systems as well as at the limit of an infinite number of particles
[22-27].
The reason is as follows.
Whereas the depleted fraction is, as mentioned above,
always zero at the infinite-particle-number limit,
even a tiny amount of depleted particles
is sufficient to completely alter the properties of variances.
Some intriguing consequences for rotating bosons
are put forward and explored in this work.

\section{Breaking up of the densities and building up of correlations in position and momentum spaces}\label{Result}

\begin{figure}[!]
\begin{center}
\includegraphics[width=0.645\columnwidth,angle=0]{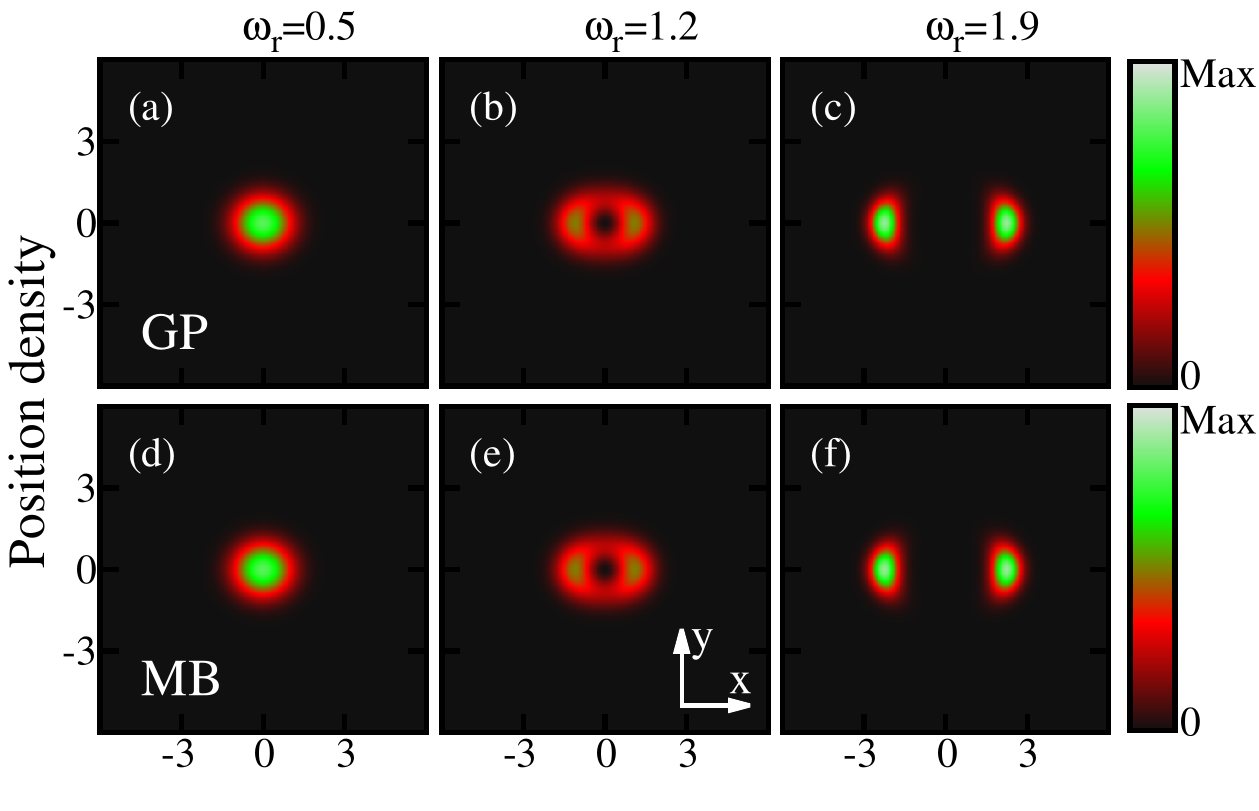}
\end{center}
\caption{Position density per particle $\frac{1}{N}\rho(\r)$ of $N=10^6$ rotating bosons
in a two-dimensional anharmonic--anisotropic trap
for three rotation frequencies $\omega_r$.
The interaction parameter is $\Lambda=0.1$.
Top row, panels (a)-(c): Mean-field results, i.e., Gross-Pitaevskii theory ($M=1$ self-consistent orbitals).
Bottom row, panels (d)-(f): Many-body results for $M=2$ self-consistent orbitals.
The respective position densities per particle are indistinguishable.
Conversely,
the anisotropies of the many-body and mean-field position variances per particle along the $x$ and $y$ directions 
become opposite with increasing rotation frequency.
See Fig.~\ref{F5} and the text for further details.
The quantities shown are dimensionless.}
\label{F1}
\end{figure}

\begin{figure}[!]
\begin{center}
\includegraphics[width=0.645\columnwidth,angle=0]{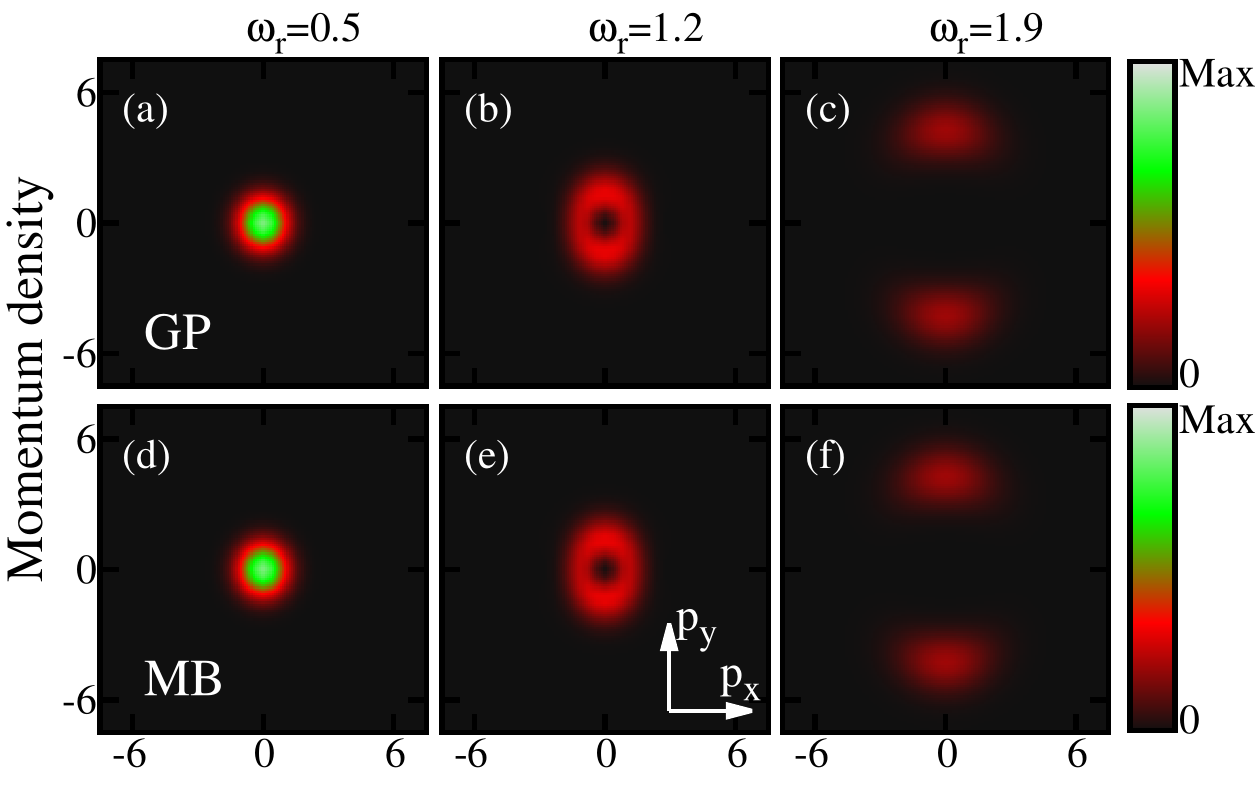}
\end{center}
\caption{Momentum density per particle $\frac{1}{N}\tilde\rho(\p)$ of $N=10^6$ rotating bosons
in a two-dimensional anharmonic--anisotropic trap
for three rotation frequencies $\omega_r$.
The interaction parameter is $\Lambda=0.1$.
Top row, panels (a)-(c): Mean-field results, i.e., Gross-Pitaevskii theory ($M=1$ self-consistent orbitals).
Bottom row, panels (d)-(f): Many-body results for $M=2$ self-consistent orbitals.
The respective momentum densities per particle are indistinguishable.
Contrarily,
the anisotropies of the many-body and mean-field momentum variances per particle
along the $p_y$ and $p_x$ directions 
become opposite with increasing rotation frequency.
See Fig.~\ref{F6} and the text for further details.
The quantities shown are dimensionless.}
\label{F2}
\end{figure}

Consider $N$ weakly-interacting repulsive bosons in a rotating anharmonic--anisotropic trap potential.
The frequency of rotation $\omega_r$ 
is increased and we follow the changes in the ground state.
The interaction parameter $\Lambda=\lambda_0(N-1)$ is held fixed and the number of bosons is increased
towards the infinite-particle-number limit, where $N=10,\ldots,10^6$.
We find as detailed below
saturation of the quantities under investigation with $N$ for a given rotation frequency $\omega_r$,
thereby providing strong numerical support of the conclusions that unique correlations in position and momentum space
exist at the limit of an infinite number of particles.

\begin{figure}[!]
\begin{center}
\includegraphics[width=0.545\columnwidth,angle=0]{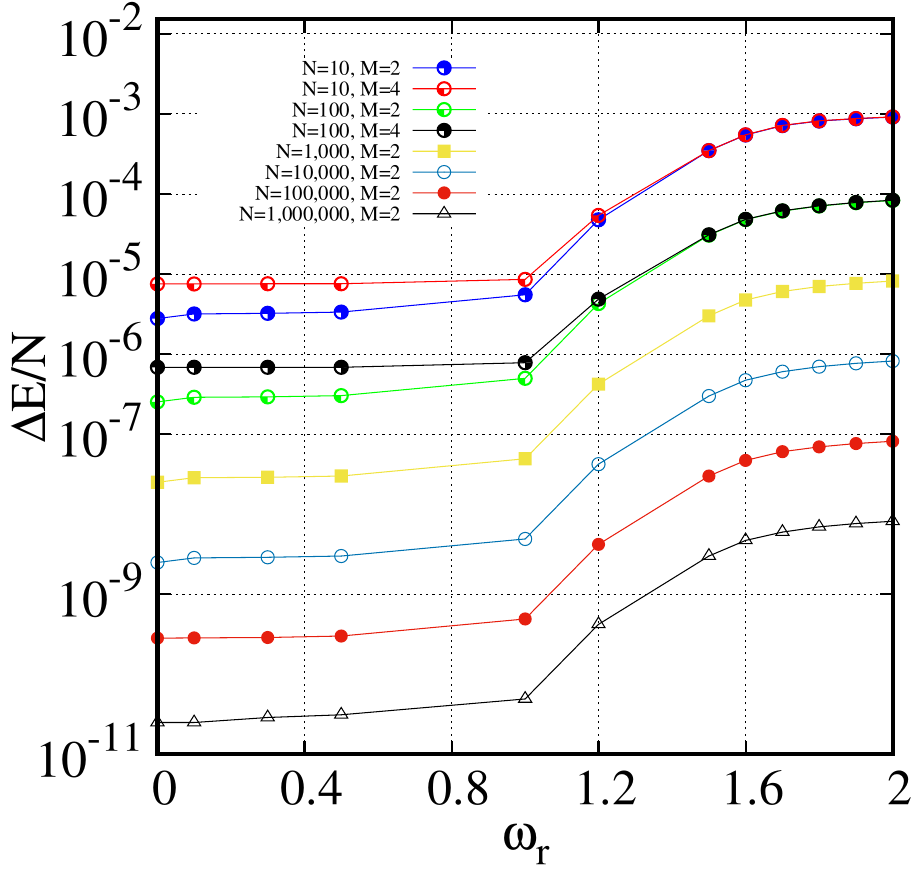}
\end{center}
\caption{Accumulation of the many-body to the mean-field energy per particle.
Depicted is the difference between the mean-field
and many-body energies per particle for $N=10,\ldots,10^6$ bosons
as a function of the rotation frequency $\omega_r$.
The interaction parameter is held fixed, $\Lambda=\lambda_0(N-1)=0.1$.
The energy difference decreases with the number of particles $N$ for a given rotation frequency $\omega_r$.
Convergence with the number of self-consistent orbitals 
is demonstrated.
See the text for further details.
Actual data is for $\omega_r=0,0.1,0.3,0.5,1.0,1.2,1.5,1.6,1.7,1.8,1.9,2.0$
and marked by symbols, the continuous curves are to guide the eye only.
The quantities shown are dimensionless.
}
\label{F3}
\end{figure}

We investigate the ground state of rotating bosons described by the 
many-particle Hamiltonian
$\hat H(\r_1,\ldots,\r_N) = \sum_{j=1}^N \hat h(\r_j) + \lambda_0 \sum_{1\le j < k}^N\hat W(\r_j-\r_k)$,
and work in the rotating frame.
The one-particle Hamiltonian is $\hat h(\r) = -\frac{1}{2}\left(\frac{\partial^2}{\partial x^2} + \frac{\partial^2}{\partial y^2}\right) 
- \omega_r \left(x \frac{1}{i}\frac{\partial}{\partial y} - y \frac{1}{i}\frac{\partial}{\partial x}\right) +
\frac{1}{4} \left(0.8 x^2 + y^2\right)^2$.
As can be seen,
the trapping potential is anharmonic and anisotropic,
being slightly wider along the $x$ direction.
The boson-boson interaction is taken to be Gaussian,
$\lambda_0 \hat W(\r-\r') = \frac{\lambda_0}{2\pi\sigma^2}e^{-\frac{\left(\r-\r'\right)^2}{2\pi\sigma^2}}$,
with width of $\sigma=0.25$.
Throughout this work the interaction parameter is $\Lambda=\lambda_0(N-1)=0.1$ and
$\hbar=1, m=1$.

The ground-state of the trapped rotating bosons is computed within the
multiconfigurational time-dependent Hartree for bosons (MCTDHB) method
[28-30].
We use the numerical implementation in MCTDH-X,
the multiconfigurational time-dependent Hartree method for indistinguishable particles software
[31-34].
The method is well documented and expanded in the literature, see, e.g.,
Refs.~[35-46].
Shortly, in the MCTDHB method
the ground state is obtained self-consistently by imaginary-time propagation
of the equations-of-motion
and determined using the variational principle.
The wavefunction is expanded by all permanents,
where $N$ bosons are distributed over $M$ self-consistent orbitals,
with optimized coefficients,
such that the energy of the ground state is minimized.
Many-body results for $M=2$ and $M=4$ orbitals are included,
where the latter are used to show convergence of the former.
Furthermore, obtaining convergence using an explicit number $M$ of self-consistent orbitals for a given number of bosons $N$
implies convergence for a larger number of bosons and
the same interaction parameter $\Lambda$.
For $M=1$ orbitals MCTDHB boils down to Gross-Pitaevskii theory \cite{GP1,GP2}.
The Hamiltonian of the rotating bosons is represented by an equidistant grid of $128 \times 128$ points
in a box of size $[-8,8) \times [-8,8)$ with periodic boundary conditions.

\begin{figure}[!]
\begin{center}
\includegraphics[width=0.545\columnwidth,angle=0]{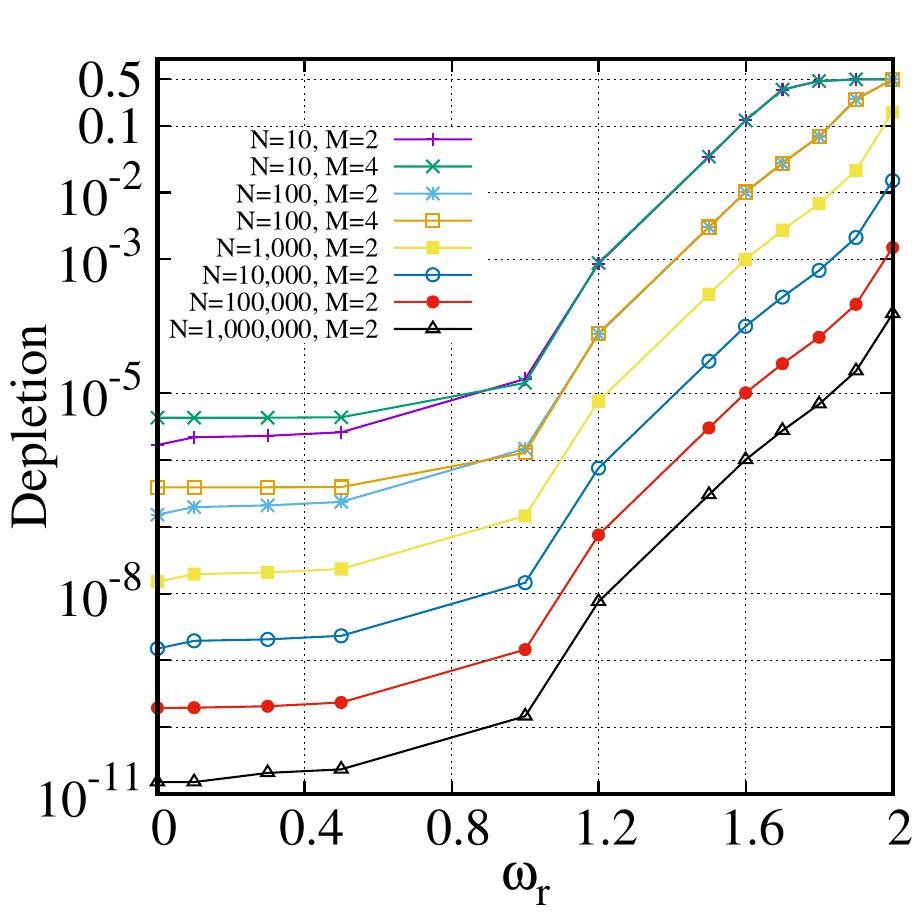}
\end{center}
\caption{Diminishing of the depleted fraction.
Shown is the depletion per particle $1-\frac{n_1}{N}$ 
for $N=10,\ldots,10^6$ bosons as a function of the rotation frequency $\omega_r$.
The interaction parameter is held fixed, $\Lambda=\lambda_0(N-1)=0.1$.
The depleted fraction decreases with the number of particles $N$ for a given rotation frequency $\omega_r$.
Convergence with the number of self-consistent orbitals 
is demonstrated.
See the text for further details.
Actual data is for $\omega_r=0,0.1,0.3,0.5,1.0,1.2,1.5,1.6,1.7,1.8,1.9,2.0$
and marked by symbols, the continuous curves are to guide the eye only.
The quantities shown are dimensionless.
}
\label{F4}
\end{figure}

\begin{figure}[!]
\begin{center}
\includegraphics[width=0.545\columnwidth,angle=0]{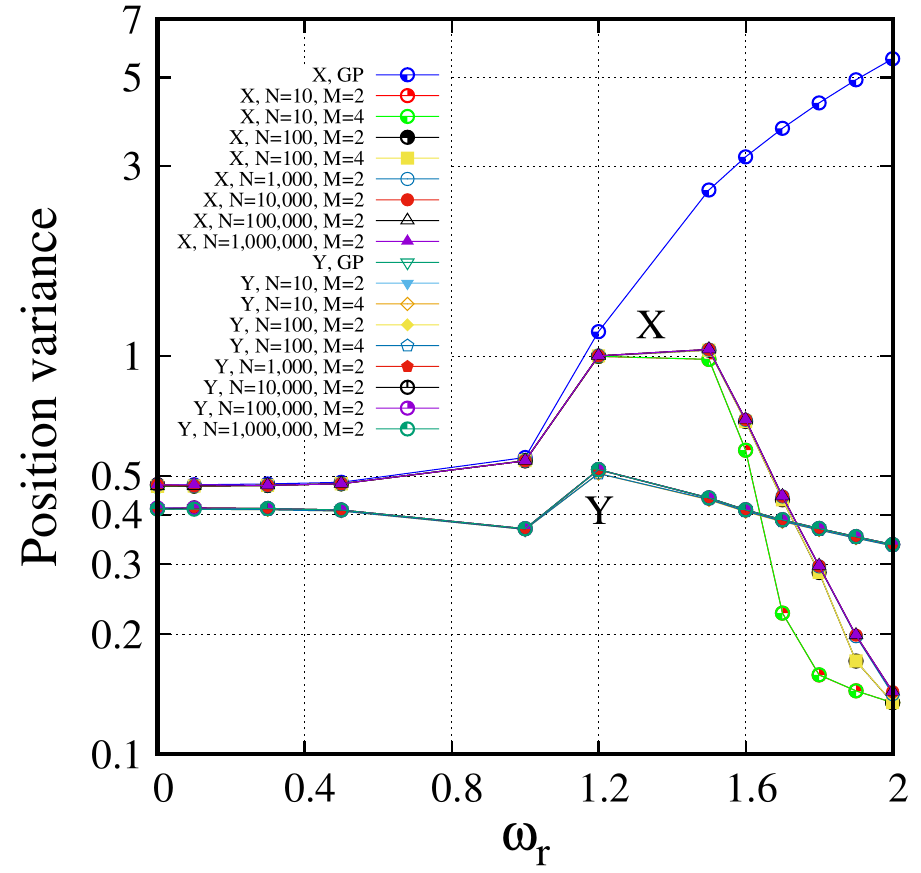}
\end{center}
\caption{Anisotropy of the position variances.
Depicted is the many-particle position variances per particle
along the $x$ and $y$ directions,
$\frac{1}{N}\Delta_{\hat X}^2$ and $\frac{1}{N}\Delta_{\hat Y}^2$,
for $N=10,\ldots,10^6$ bosons as a function of the rotation frequency $\omega_r$.
The interaction parameter is $\Lambda=\lambda_0(N-1)=0.1$.
The anisotropies of the many-particle position variances
at the many-body and mean-field levels of theory 
become opposite to each other at rotation frequencies above about $\omega_r=1.75$.
Convergence with the number of self-consistent orbitals 
is demonstrated.
See the text for further details.
Actual data is for $\omega_r=0,0.1,0.3,0.5,1.0,1.2,1.5,1.6,1.7,1.8,1.9,2.0$
and marked by symbols, the continuous curves are to guide the eye only.
The quantities shown are dimensionless.
}
\label{F5}
\end{figure}

\begin{figure}[!]
\begin{center}
\includegraphics[width=0.545\columnwidth,angle=0]{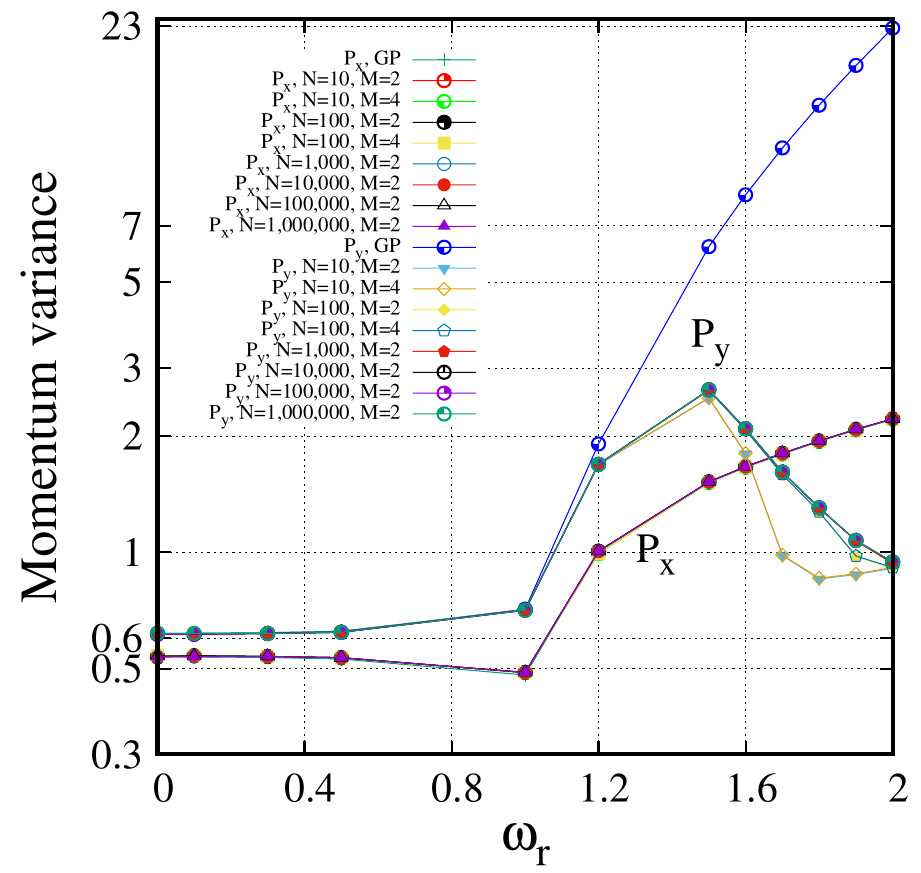}
\end{center}
\caption{Anisotropy of the momentum variances.
Shown is the many-particle momentum variances per particle
along the $p_x$ and $p_y$ directions,
$\frac{1}{N}\Delta_{\hat P_X}^2$ and $\frac{1}{N}\Delta_{\hat P_Y}^2$,
for $N=10,\ldots,10^6$ bosons as a function of the rotation frequency $\omega_r$.
The interaction parameter is $\Lambda=\lambda_0(N-1)=0.1$.
The anisotropies of the many-particle momentum variances
at the many-body and mean-field levels of theory 
become opposite to each other at rotation frequencies above about $\omega_r=1.7$.
Convergence with the number of self-consistent orbitals 
is demonstrated.
See the text for further details.
Actual data is for $\omega_r=0,0.1,0.3,0.5,1.0,1.2,1.5,1.6,1.7,1.8,1.9,2.0$
and marked by symbols, the continuous curves are to guide the eye only.
The quantities shown are dimensionless.
}
\label{F6}
\end{figure}

Fig.~\ref{F1} displays the position densities per particle, $\frac{1}{N}\rho(\r)$, of $N=10^6$ bosons
computed at the $M=2$ many-body and $M=1$ mean-field levels of theory,
and likewise Fig.~\ref{F2} is for the momentum densities per particle, $\frac{1}{N}\tilde\rho(\p)$.
The respective densities per particle cannot be distinguished from each other,
in accordance with the mathematical-rigorous literature result
[19-21].
Importantly, with increasing of the rotation frequency
the density splits
into two parts both in position and in momentum spaces.
In position space, due to the growth of
a centrifugal barrier the density splits along the $x$ direction; recall that the anharmonic--anisotropic trap potential is slightly wider along the $x$ axis.
Side by side, the momentum density splits along the $p_y$
direction due to the rotation
which imprints linear momentum in the orthogonal direction(s) on the spatially-split bosons.
The accuracy and adequacy of the $M=2$ computations for the
two-dimensional trap and various rotation frequencies considered here
are verified against $M=4$ computations
with $N=10$ and $N=100$ bosons,
see Figs.~\ref{F3}-\ref{F6} and discussions below. 

We proceed and now examine other quantities and their values as the infinite-particle-number limit is taken.
Fig.~\ref{F3} depicts the difference between the mean-field and many-body
ground-state energies per particle, $\frac{\Delta E}{N}$, as a function of the rotation frequency, 
for the numbers of bosons $N=10,\ldots,10^6$, and at a constant interaction parameter $\Lambda=0.1$.
The difference is found to increase with $\omega_r$ for a given $N$
and to decrease with $N$ for a given $\omega_r$.
The latter is in accordance with the mathematical-rigorous result in the literature
[19-21].
To verify that $M=2$ self-consistent orbitals accurately describe the difference in the energies per particle $\frac{\Delta E}{N}$
in the two-dimensional anharmonic--anisotropic trap for the various rotation frequencies,
we have also performed calculations using $M=4$ self-consistent orbitals for $N=10$ and $N=100$ bosons.
The results
are seen to fall on top of the respective $M=2$ results to within $10^{-4}$ ($10^{-5}$) for $N=10$ ($N=100$) bosons,
see Fig.~\ref{F3}.
Since the interaction parameter is kept fixed,
this implies the accuracy of the $M=2$ and $N>100$ results as well.

Fig.~\ref{F4} presents the depletion per particle, $1-\frac{n_1}{N}$.
Here, $n_1$ is the largest occupation number 
obtained from diagonalization of the reduced one-particle 
density matrix
[49-51]
$\rho^{(1)}(\r,\r')=N\int d\r_2 \ldots d\r_N \Psi^\ast(\r',\r_2,\ldots,\r_N)\Psi(\r,\r_2,\ldots,\r_N)$,
where $\Psi(\r_1,\r_2,\ldots,\r_N)$ is normalized to unity.
We find that the depletion fraction increases with the rotation frequency $\omega_r$.
In fact, for $N=10$ bosons the system becomes two-fold fragmented \cite{BUPMB}
and as can be seen in Fig.~\ref{F4} also for $N=100$ bosons the system becomes two-fold fragmented,
however for relatively faster rotations.
Fragmentation means the macroscopic occupation of more
than one eigenvalue of the reduced one-particle density matrix, see, e.g.,
Refs.~[52-60].
Increasing the number of particles more and for a given rotation frequency,
the depleted fraction is seen to diminish further and further {\it en route} to zero depletion,
which is in accordance with the literature
[19-21].
To corroborate that $M=2$ self-consistent orbitals accurately describe the depleted fraction $1-\frac{n_1}{N}$
in the two-dimensional anharmonic--anisotropic trap for the various rotation frequencies $\omega_r$ considered,
we have also performed calculations using $M=4$ self-consistent orbitals for $N=10$ and $N=100$ bosons.
The latter are 
seen to fall on top of the respective $M=2$ results to better than $10^{-4}$ ($10^{-5}$) for $N=10$ ($N=100$) bosons,
see Fig.~\ref{F4}.
Since the interaction parameter is kept fixed,
this implies the accuracy of the $N>100$ results with $M=2$ self-consistent orbitals as well.

Let us summarize our results so far.
The rotation leads to splitting of the bosons into two sub-clouds
in position space along the $x$ direction,
see Fig.~\ref{F1},
and into two sub-clouds in momentum space along the $p_y$ direction,
see Fig.~\ref{F2}.
This dual behavior of the density suggests that
two effective double wells are formed,
one in position space and the other one in momentum space.
The ramifications will be exploited below.
We also found, see Fig.~\ref{F4}, that the rotation leads to fragmentation for relatively small numbers of particles,
explicitly for $N=10$ \cite{BUPMB} and $N=100$ bosons in the examples considered.
As the number of bosons is increased while keeping the interaction parameter $\Lambda$ fixed,
the fragmentation and later on the depletion decrease more and more.
Additionally, Figs.~\ref{F1}, \ref{F2}, \ref{F3}, and \ref{F4} have demonstrated numerically
the literature results on the density per particle, energy per particle, and depletion per particle
of a trapped rotating bosonic system in two spatial dimensions
with a constant interaction parameter
at the infinite-particle-number limit
[19-21].

We now move to investigate the many-particle position and momentum variances of the interacting bosons in the rotating system,
which would lead to one of the main results of the present work.
Complementary results on the many-particle angular-momentum variance
and average angular momentum are collected and briefly discussed in \ref{APP}. 
Consider a many-particle observable $\hat O=\sum_{j=1}^N \hat o(\r_j)$.
In the present work we examine the position, $\hat X$, $\hat Y$,
momentum $\hat P_X$, $\hat P_Y$,
and angular-momentum $\hat L_Z$ many-particle operators.
The many-particle variance per particle is given by
$\frac{1}{N}\Delta^2_{\hat O} = \frac{1}{N}\left[\langle\Psi|\hat O^2|\Psi\rangle - \langle\Psi|\hat O|\Psi\rangle^2\right]$,
where
$\frac{1}{N}{\langle\Psi|\hat O|\Psi\rangle}$ is the average per particle.
Intricacies arise for the variance because $\hat O^2 = \sum_{j=1}^N \hat o^2(\r_j) + 2\sum_{1\le j < k}^N \hat o(\r_j)\hat o(\r_k)$
contains both one-particle and two-particle operators \cite{VAR0}.

Figs.~\ref{F5} and \ref{F6} present the results.
It is instrumental to analyze them in view of
the study of bosons in a two-dimensional double-well potential,
without rotation of course,
at the infinite-particle-number limit, Ref.~\cite{ANINF}.
We begin with position space.
At the mean-field level of theory,
the variance per particle along the $x$ direction, $\frac{1}{N}\Delta_{\hat X}^2$,
grows monotonously with the rotation frequency,
in accordance with the overall broadening of
the density, see Fig.~\ref{F1}.
The position density hardly changes its width along the $y$ direction,
and, accordingly, $\frac{1}{N}\Delta_{\hat Y}^2$, does not change much, see Fig.~\ref{F5}.
At the many-body level of theory,
$\frac{1}{N}\Delta_{\hat X}^2$ initially follows the mean-field variance for slow rotations and then they begin to depart from each other,
and $\frac{1}{N}\Delta_{\hat X}^2$ starts to decrease as $\omega_r$ further increases.
This behavior of the many-particle position variance has been found in a double-well potential as a function of the barrier height \cite{VAR0, ANINF}
and now seen to occur in the barrierless anharmonic--anisotropic potential under rotation \cite{BUPMB}.
This is a many-body signature that the rotation leads to an effective double well along the $x$ direction.
In this respect, the splitting of the density in Fig.~\ref{F1} is a mean-field signature
that the rotation generates an effective double well, see \cite{BUPMF}.
Finally, the many-body variance along the $y$ direction $\frac{1}{N}\Delta_{\hat Y}^2$ essentially coincides with the
mean-field result, see Fig.~\ref{F5}, similarly to the double-well case \cite{ANINF}.

Examining now jointly $\frac{1}{N}\Delta_{\hat X}^2$ and $\frac{1}{N}\Delta_{\hat Y}^2$,
both at the many-body and mean-field levels,
we find the anisotropy of the position variances \cite{ANICP,ANINF}.
Namely, up to a certain parameter, here the rotation frequency of about $w_r=1.75$,
the many-particle position variances satisfy $\frac{1}{N}\Delta_{\hat X}^2 > \frac{1}{N}\Delta_{\hat Y}^2$
both at the mean-field and many-body levels of theory,
i.e., the bosons possess the same mean-field and many-body position anisotropies. 
Above about $w_r=1.75$ the many-particle
position variances continue to satisfy $\frac{1}{N}\Delta_{\hat X}^2 > \frac{1}{N}\Delta_{\hat Y}^2$
at the mean-field level but now obey $\frac{1}{N}\Delta_{\hat X}^2 < \frac{1}{N}\Delta_{\hat Y}^2$ at the many-body level,
namely,
the bosons possess opposite mean-field and many-body position anisotropies,
see Fig.~\ref{F5}. 
This conclusion is demonstrated to be valid at the limit of an infinite number of particles,
as the curves for $N=10,\ldots,10^6$ bosons at a fixed interaction parameter $\Lambda$ quickly
accumulate on top of each other.
Furthermore, the convergence is verified by the coincidence
of respective $N=10$ and $N=100$ results with $M=2$ and $M=4$ self-consistent orbitals.  
We have thus generalized and extended the result for the anisotropy of the position variance obtained
for rotating fragmented bosons in \cite{BUPMB} to the infinite-particle-number limit
where the bosons are $100\%$ condensed.

Last but not least, we proceed to analyze the variances in momentum space, see Fig.~\ref{F6},
which, as can be anticipated by comparing the respective densities in Figs.~\ref{F1} and \ref{F2},
would exhibit complementary and analogous behavior to the position variances.
At the mean-field level of theory,
the variance per particle along the $p_y$ axis, $\frac{1}{N}\Delta_{\hat P_Y}^2$,
grows monotonously with $\omega_r$,
in accordance with the overall widening of
the momentum density, see Fig.~\ref{F2}.
The width of the momentum density along the $p_x$ direction changes rather mildly,
and, correspondingly, $\frac{1}{N}\Delta_{\hat P_X}^2$ grows slower than $\frac{1}{N}\Delta_{\hat P_Y}^2$, see Fig.~\ref{F6}.
At the many-body level of theory,
$\frac{1}{N}\Delta_{\hat P_Y}^2$ follows the mean-field variance for slow rotations and then they start to depart from each other,
following by a complete change of behavior, namely, decreasing values, with increasing rotation frequencies.
This dependence of the many-particle momentum variance is unique to the
barrierless anharmonic--anisotropic potential under rotation \cite{BUPMB}.
As far as we can tell,
it has no analog and does not occur in a double-well potential as a function of the barrier height in any dimension \cite{VAR0, ANINF}.
In a sense,
this is a many-body fingerprint that the rotation leads to an effective double well along the $p_y$ direction.
In this regard, the splitting of the density in Fig.~\ref{F2} is a mean-field signature
that the rotation generates an effective double well in momentum space.
Finally, the many-body momentum variance along the $p_x$ direction, $\frac{1}{N}\Delta_{\hat P_X}^2$, practically coincides with the
mean-field result, see Fig.~\ref{F5}, which reminds one
of the behavior of the momentum variance in the orthogonal direction
in the double-well case \cite{ANINF}.

Inspecting now $\frac{1}{N}\Delta_{\hat P_Y}^2$ and $\frac{1}{N}\Delta_{\hat P_X}^2$ together,
both at the many-body and mean-field levels of theory,
we find the anisotropy of the momentum variances.
Namely, up to a certain parameter, here the rotation frequency of about $w_r=1.7$,
the many-particle momentum variances satisfy $\frac{1}{N}\Delta_{\hat P_Y}^2 > \frac{1}{N}\Delta_{\hat P_X}^2$
both at the mean-field and many-body levels of theory,
namely, the bosons possess the same mean-field and many-body momentum anisotropies. 
For rotation frequencies above about $w_r=1.7$,
the relation
$\frac{1}{N}\Delta_{\hat P_Y}^2 > \frac{1}{N}\Delta_{\hat P_X}^2$ continues to hold
for the mean-field quantities,
however the many-body variances satisfy 
$\frac{1}{N}\Delta_{\hat P_Y}^2 < \frac{1}{N}\Delta_{\hat P_X}^2$,
meaning that
the bosons possess opposite mean-field and many-body momentum anisotropies,
see Fig.~\ref{F6}. 
This finding is shown to be valid at the infinite-particle-number limit
since the curves for $N=10,\ldots,10^6$ bosons at fixed $\Lambda$ quickly
gather on top of each other.
Moreover, the convergence is confirmed by the coincidence
of corresponding $N=10$ and $N=100$ results with $M=2$ and $M=4$ self-consistent orbitals.  
We have thus generalized and extended the result for the anisotropy of the momentum variance put forward
for rotating fragmented bosons in \cite{BUPMB} to the limit of an infinite number of particles
in which the bosons are $100\%$ condensed.

A final comment.
So far, the anisotropy of the many-particle momentum variance
has only been found at the infinite-particle-number limit 
in the out-of-equilibrium dynamics of a tapped Bose-Einstein condensate 
\cite{ANICP}, also see \cite{MORPH}.
In Ref.~\cite{ANINF} it has been questioned whether a ground state
whose momentum variance exhibits opposite
anisotropy at the limit of an infinite number of particles can be found.
The present work provides in our opinion an aesthetic two-in-one answer,
that rotating bosons in a barrierless anharmonic--anisotropic trap
can possess opposite anisotropies of both the position and momentum many-particle variances 
in their ground state simultaneously,
while being fully condensed
at the infinite-particle-number limit.
 
\section{Concluding remarks}\label{Sum}

The ground state of a rotating Bose-Einstein condensate in a two-dimensional anharmonic--anisotropic trap potential
is analyzed numerically at the infinite-particle-number limit.
First, we show that the density in position space splits in two along the $x$ direction,
and, side by side, the density in momentum space splits in two along the $p_y$ direction.
The resulting unusually-split bosonic cloud can thus be interpreted as living
in effective double-well potentials both in position and momentum spaces. 
Furthermore, the bosons exhibit unique correlations.
To this end,
it is demonstrated that the anisotropies of the many-particle 
position and momentum variances become opposite when computed
at the many-body and mean-field levels of theory with increasing of
the rotation frequency,
despite the system being $100\%$ condensed.
It would be interesting to explore generalizations, for other trap geometries, in three spatial dimensions,
and to apply the present findings to the out-of-equilibrium dynamics of rotating bosons.

\ack

This work was supported by the Israel Science Foundation (Grant No.~1516/19).
Computation time at the High Performance Computing Center
Stuttgart (HLRS) is gratefully acknowledged.

\appendix\section{Angular-momentum properties}\label{APP}

\begin{figure}[!]
\begin{center}
\hglue -0.5 truecm
\includegraphics[width=0.445\columnwidth,angle=0]{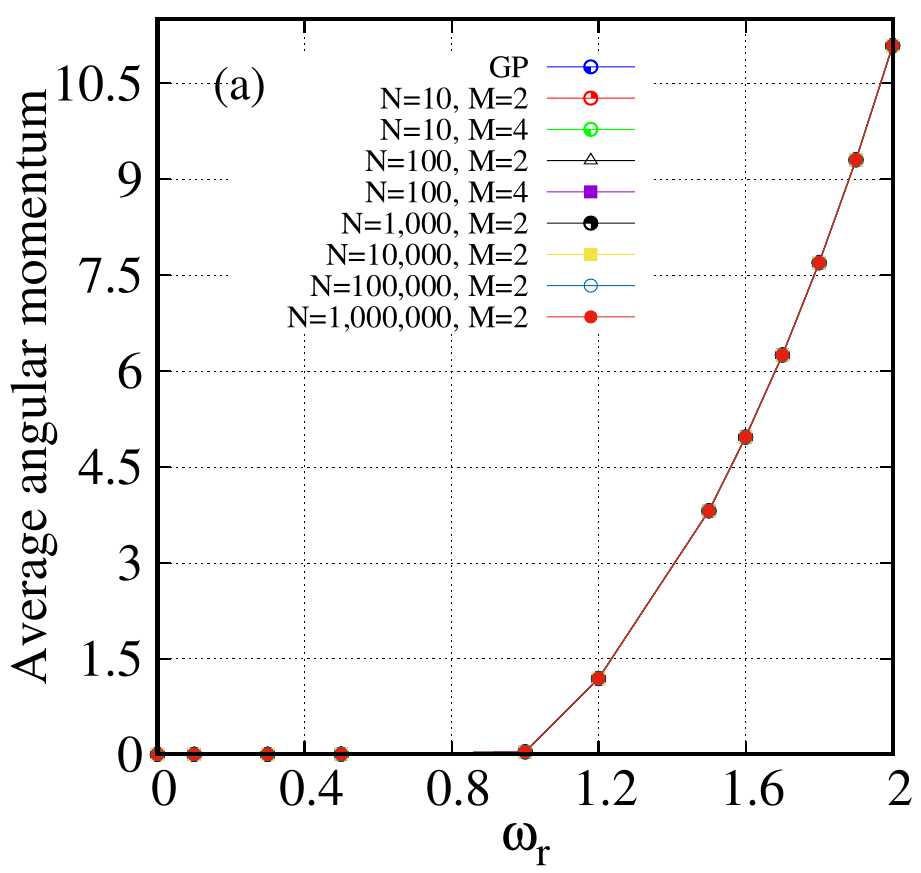}
\hglue 0.5 truecm
\includegraphics[width=0.445\columnwidth,angle=0]{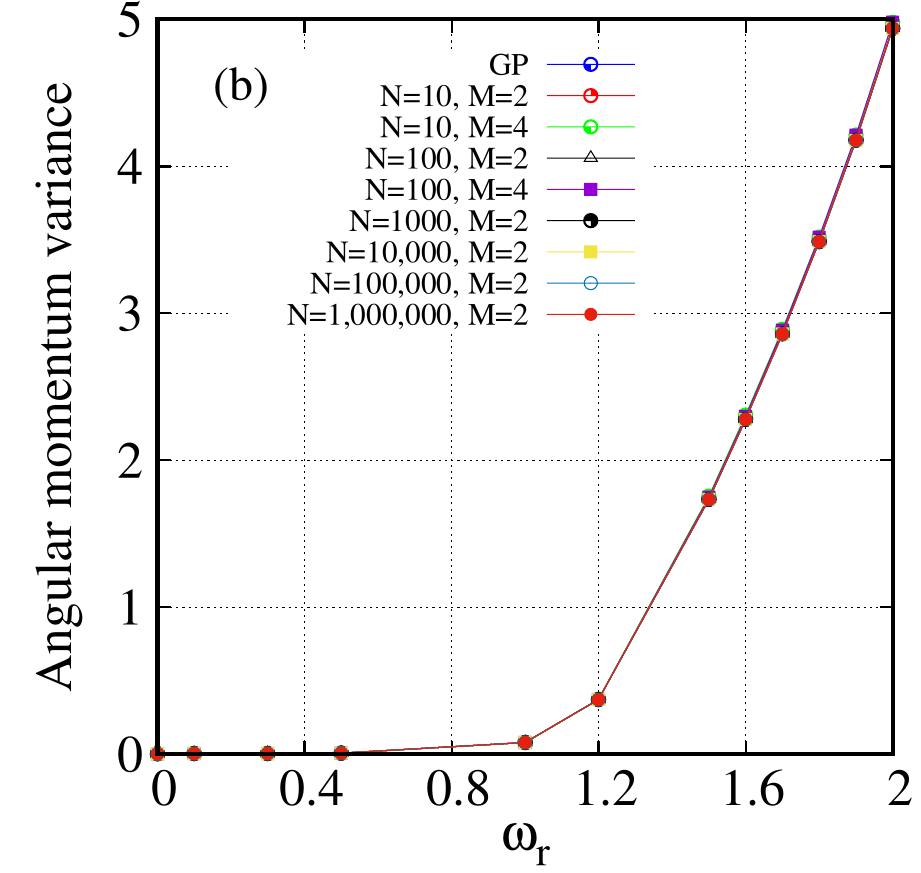}
\end{center}
\caption{Angular-momentum properties.
Shown is (a) the average angular-momentum per particle, $\frac{1}{N}\langle\Psi|\hat L_Z|\Psi\rangle$,
and (b) the many-particle angular-momentum variance per particle, $\frac{1}{N}\Delta_{\hat L_Z}^2$,
for $N=10,\ldots,10^6$ bosons as a function of the rotation frequency $\omega_r$.
The interaction parameter is $\Lambda=\lambda_0(N-1)=0.1$.
The bosons acquire with increasing rotation average angular momentum and fluctuations.
Interestingly and unlike the many-particle position and momentum variances,
see Figs.~\ref{F5} and ~\ref{F6},
the many-body and mean-field results are nearly the same.
The numbers $M$ of self-consistent orbitals are indicated in the panels.
Convergence with $M$ is demonstrated, of course, for the average angular momentum.
One would require $M=8$ self-consistent orbitals \cite{BUPMB}
to show that $M=4$ self-consistent orbitals suffice for convergence of
the angular-momentum variance.
See the text for further details.
Actual data is for $\omega_r=0,0.1,0.3,0.5,1.0,1.2,1.5,1.6,1.7,1.8,1.9,2.0$
and marked by symbols, the continuous curves are to guide the eye only.
The quantities shown are dimensionless.
}
\label{F7}
\end{figure}

The appendix presents additional analysis, of angular-momentum properties.
Fig.~\ref{F7} depicts 
the average angular-momentum per particle, $\frac{1}{N}\langle\Psi|\hat L_Z|\Psi\rangle$,
and the many-particle angular-momentum variance per particle, $\frac{1}{N}\Delta_{\hat L_Z}^2$.
The system is slightly anisotropic and hence not rotationally symmetric,
but for no rotation it is reflection symmetric.
For weak rotations, there is hardly any deformation of the density, see Figs.~\ref{F1} and \ref{F2},
and, side by side, the average angular momentum and its variance are practically zero, see Fig.~\ref{F7}.
With increasing rotation, the bosons absorb angular momentum
while the density distorts in position and momentum spaces.
Clearly, the system is not an eigenfunction of the many-particle angular-momentum operator
and its variance increases further and further with the rotation too. 

Comparing the mean-field and many-body results one gets, of course,
no difference between the expectation values per particle in the limit of an infinite number of particles,
as is expected from theory.
Interestingly,
very little differences are found between the mean-field and many-body variances $\frac{1}{N}\Delta_{\hat L_Z}^2$,
unlike the above results for the respective position and momentum quantities.
We attribute this situation to the geometry of the system, but further analysis is needed to prove that, also see \cite{ANALYSIS}.
Perhaps, subjecting the bosons to an artificial gauge field
[61-64]
could generate larger differences
for the angular-momentum variances.
This is a good place to conclude
the present investigations.

\end{document}